\documentclass[prb,twocolumn,showpacs,preprintnumbers,amsmath,aps]{revtex4-1}
\usepackage{graphicx,hyperref}% Include figure files
\usepackage{bm}% bold math
\usepackage{subfigure}% bold math

\newcommand\tr{\text{Tr}}

\def\nbC{{\mathchoice {\setbox0=\hbox{$\displaystyle\rm C$}%
\hbox{\hbox to0pt{\kern0.4\wd0\vrule height0.9\ht0\hss}\box0}}
{\setbox0=\hbox{$\textstyle\rm C$}\hbox{\hbox
to0pt{\kern0.4\wd0\vrule height0.9\ht0\hss}\box0}}
{\setbox0=\hbox{$\scriptstyle\rm C$}\hbox{\hbox
to0pt{\kern0.4\wd0\vrule height0.9\ht0\hss}\box0}}
{\setbox0=\hbox{$\scriptscriptstyle\rm C$}\hbox{\hbox
to0pt{\kern0.4\wd0\vrule height0.9\ht0\hss}\box0}}}}
\def\nbQ{{\mathchoice {\setbox0=\hbox{$\displaystyle\rm
Q$}\hbox{\raise
0.15\ht0\hbox to0pt{\kern0.4\wd0\vrule height0.8\ht0\hss}\box0}}
{\setbox0=\hbox{$\textstyle\rm Q$}\hbox{\raise
0.15\ht0\hbox to0pt{\kern0.4\wd0\vrule height0.8\ht0\hss}\box0}}
{\setbox0=\hbox{$\scriptstyle\rm Q$}\hbox{\raise
0.15\ht0\hbox to0pt{\kern0.4\wd0\vrule height0.7\ht0\hss}\box0}}
{\setbox0=\hbox{$\scriptscriptstyle\rm Q$}\hbox{\raise
0.15\ht0\hbox to0pt{\kern0.4\wd0\vrule height0.7\ht0\hss}\box0}}}}
\def\nbT{{\mathchoice {\setbox0=\hbox{$\displaystyle\rm
T$}\hbox{\hbox to0pt{\kern0.3\wd0\vrule height0.9\ht0\hss}\box0}}
{\setbox0=\hbox{$\textstyle\rm T$}\hbox{\hbox
to0pt{\kern0.3\wd0\vrule height0.9\ht0\hss}\box0}}
{\setbox0=\hbox{$\scriptstyle\rm T$}\hbox{\hbox
to0pt{\kern0.3\wd0\vrule height0.9\ht0\hss}\box0}}
{\setbox0=\hbox{$\scriptscriptstyle\rm T$}\hbox{\hbox
to0pt{\kern0.3\wd0\vrule height0.9\ht0\hss}\box0}}}}
\def\nbS{{\mathchoice
{\setbox0=\hbox{$\displaystyle     \rm S$}\hbox{\raise0.5\ht0%
\hbox to0pt{\kern0.35\wd0\vrule height0.45\ht0\hss}\hbox
to0pt{\kern0.55\wd0\vrule height0.5\ht0\hss}\box0}}
{\setbox0=\hbox{$\textstyle        \rm S$}\hbox{\raise0.5\ht0%
\hbox to0pt{\kern0.35\wd0\vrule height0.45\ht0\hss}\hbox
to0pt{\kern0.55\wd0\vrule height0.5\ht0\hss}\box0}}
{\setbox0=\hbox{$\scriptstyle      \rm S$}\hbox{\raise0.5\ht0%
\hboxto0pt{\kern0.35\wd0\vrule height0.45\ht0\hss}\raise0.05\ht0%
\hbox to0pt{\kern0.5\wd0\vrule height0.45\ht0\hss}\box0}}
{\setbox0=\hbox{$\scriptscriptstyle\rm S$}\hbox{\raise0.5\ht0%
\hboxto0pt{\kern0.4\wd0\vrule height0.45\ht0\hss}\raise0.05\ht0%
\hbox to0pt{\kern0.55\wd0\vrule height0.45\ht0\hss}\box0}}}}
\def\nbZ{{\mathchoice {\hbox{$\sf\textstyle Z\kern-0.4em Z$}}
{\hbox{$\sf\textstyle Z\kern-0.4em Z$}}
{\hbox{$\sf\scriptstyle Z\kern-0.3em Z$}}
{\hbox{$\sf\scriptscriptstyle Z\kern-0.2em Z$}}}}
\begin{document}

\title{Supersymmetry and its spontaneous breaking in the random field Ising model}

\author{
Matthieu Tissier} %\email{tissier@lptl.jussieu.fr}
\affiliation{LPTMC, CNRS-UMR 7600, Universit\'e Pierre et Marie Curie,
bo\^ite 121, 4 Pl. Jussieu, 75252 Paris c\'edex 05, France}

\author{Gilles Tarjus} %\email{tarjus@lptl.jussieu.fr}
\affiliation{LPTMC, CNRS-UMR 7600, Universit\'e Pierre et Marie Curie,
bo\^ite 121, 4 Pl. Jussieu, 75252 Paris c\'edex 05, France}

%\date{\today}

\begin{abstract}
We provide a resolution of one of the long-standing puzzles in the
theory of disordered systems. By reformulating the functional
renormalization group (FRG) for the critical behavior of the random
field Ising model in a superfield formalism, we are able to follow the
associated supersymmetry and its spontaneous breaking along the FRG
flow. Breaking is shown to occur below a critical dimension
$d_{DR}\simeq 5.1$ and leads to a breakdown of the ``dimensional
reduction'' property. We compute the critical exponents as a function
of dimension and give evidence that scaling is described by three
independent exponents.
\end{abstract}


\maketitle

The random field Ising model (RFIM) is one of the archetypal disordered systems\cite{nattermann98} and, ever since the seminal work of Imry and Ma,\cite{imry-ma75}, its long-distance properties have provided puzzles that largely remain pending. About thirty years ago, Parisi and Sourlas in a beautiful $2$-page letter \cite{parisi79} related the critical behavior of an Ising ferromagnet coupled to a random magnetic field to a supersymmetric
scalar field theory and showed that the supersymmetry (SUSY) leads to a ``dimensional reduction'' (DR) property by which the behavior of the RFIM in $d$ dimensions is identical to that of the system without disorder in $d-2$. It is well established that the SUSY construction for the RFIM and the associated DR property actually break down in low dimensions,\cite{imry-ma75,imbrie84,parisi84b} but the issue has not been satisfactorily settled. In a recent work, we showed by means of a nonperturbative functional renormalization group (NP-FRG) approach\cite{tarjus04,tissier06} that the breakdown of DR is related to the appearance of a nonanalytic dependence of the effective action in the dimensionless fields. Physically, this arises from the presence of rare collective events known as ''avalanches'' or "shocks". However, our formalism  could not at all address the question of SUSY and its breaking. The purpose of the present letter is to fill this gap and provide a complete picture of the critical behavior of the RFIM.

Our starting point is the RFIM field-theoretical description in terms of a scalar field
$\phi(x)$ in a $d$-dimensional space and a bare action $S[\phi;h]$ given by
\begin{equation}
\label{eq_ham_dis}
S= \int_{x} \bigg\{ \frac
{1}{2}  \left(\partial_{\mu} \phi(x) \right) ^2 + U_B(\phi(x)) -
h(x) \phi(x) \bigg\} ,
\end{equation}
where $ \int_{x} \equiv \int d^d x$,  $U_B(\phi)= (\tau/2) \phi^2  +  (u/4!) \phi^4$, and $h(x)$ is a
random magnetic field that is taken from a Gaussian distribution with a zero
mean and a variance $\overline{h(x)h(y)}=\Delta_B \ \delta^{(d)}(x-y)$. Taking advantage of the fact that, at long-distance, the thermal fluctuations are negligible compared to those induced by disorder (formally, the critical behavior is controlled by a zero-temperature fixed point\cite{nattermann98}), one can focus on the solution of the stochastic field equation\cite{parisi79}
\begin{equation}
\dfrac{\delta S[\phi;h]}{\delta \phi(x)}=0.
\end{equation}
Provided the solution is unique, the correlation functions of the $\phi$ field are then obtained from appropriate derivatives of a generating functional that can be built through standard field-theoretical techniques.\cite{zinnjustin89} One first introduces  auxiliary fields, a bosonic ``response'' field $\hat{\phi}(x)$ and two fermionic ``ghost'' fields  $\psi(x)$ and  $\bar{\psi}(x)$, as well as  linearly coupled sources, and one explicitly performs the average over the Gaussian disorder. After constructing a superspace by adding to the $d$-dimensional Euclidean space with coordinates $x=\left\lbrace x^\mu\right\rbrace $ two anti-commuting Grassmann coordinates $\theta,\bar{\theta}$ (satisfying $\theta^2=\bar{\theta}^2=\theta \bar{\theta}+\bar{\theta}\theta=0$), the resulting functional can be cast in a superfield formalism:\cite{parisi79}
\begin{equation}
  \label{eq_part_func}
\mathcal Z[\mathcal J]=\int\mathcal D\Phi  \exp \left(- S_{ss}[\Phi] +   \int_{\underline x}  \mathcal J(\underline x)  \Phi(\underline x)\right) ,
\end{equation}
with
\begin{equation}
S_{ss}[\Phi]=\int_{\underline{x}} \frac 12 [- \Phi(\underline{x})\Delta_{ss}\Phi(\underline{x})+U_{B}(\Phi(\underline{x}))],
\end{equation}
where we have introduced the superfield $\Phi(\underline{x})=\phi(x) + \bar{\theta} \psi(x)+ \bar{\psi}(x) \theta + \bar{\theta}\theta \hat{\phi}(x)$, the supersource $\mathcal J(\underline x)=J(x) + \bar{\theta} K(x)+ \bar{K}(x) \theta + \bar{\theta}\theta \hat{J}(x)$, and the superlaplacian $\Delta_{ss}=\partial_\mu^2+\Delta_B \partial_\theta \partial_{\bar{\theta}}$;  $\underline{x}=(x,\theta,\bar{\theta})$ denotes the coordinates in superspace and $ \int_{\underline{x}} \equiv \int d^d x d\theta d\bar{\theta}$. The $\phi$-field correlation functions are obtained by functional derivatives of $\mathcal Z[\mathcal J]$ with respect to $\hat{J}$ (evaluated for $K=\hat{K}=\hat{J}=0$).

The action $S_{ss}$ is invariant under a large group of transformations that mix bosonic and fermionic fields (hence the name, SUSY): translations and rotations in the $d$-dimensional Euclidean and $2$-dimensional Grassmannian subspaces
and ``superrotations'' that preserve the superdistance, $\underline{x}^2=x^2+ \frac{4}{\Delta_B}\theta \bar{\theta}$. These superrotations can be represented by the generators ${\mathcal Q}_\mu=x^\mu \partial_{\bar{\theta}}+ \frac{2}{\Delta_B} \theta \partial_{\mu}$ and $\bar{\mathcal Q}_\mu=-x^\mu \partial_\theta + \frac{2}{\Delta_B} \bar{\theta} \partial_{\mu}$. The presence of this SUSY, more precisely of the superrotations, was shown to lead to DR.\cite{parisi79,cardy83}  One knows, however, that the whole formal construction collapses when the stochastic field equation has more than one solution, which is the case in the region of interest.\cite{parisi84b}

We propose a resolution of the above problem that allows one to study
the SUSY and its spontaneous breaking. To this end we upgrade our
NP-FRG approach\cite{tarjus04,tissier06} to a superfield
formulation. The key points involve (i) adding an infrared (IR)
regulator that enforces a progressive account of the fluctuations of
both the $\phi$ field and the disorder while ensuring that the initial
condition of the RG flow satisfies the SUSY, (ii) considering copies
of the original disordered system, which gives access to the full
functional field dependence of the renormalized cumulants of the
disorder, (iii) enforcing that a single solution of the stochastic
field equation (the ground state) is taken into account for each copy,
and (iv) using the Ward-Takahashi (WT) identities associated with the
SUSY to ensure that neither the regulator nor the approximations
explicitly break the SUSY. We stress that the introduction of copies (or replicas)
is necessary to describe nonanalyticities in the renormalized
cumulants stemming from the occurence of ''avalanches'', even in a
superfield formalism. This, unusual, combined use of supersymmetric formalism {\em and} replicas is central for overcoming the flaws of the Parisi-Sourlas construction.

Extending our previous work to the superfield theory, we introduce a generating functional of the correlation functions at the running scale $k$ for an arbitrary number $n$ of copies of the system (coupled to the same random field but submitted to different external sources),
\begin{equation}
 \begin{aligned}
 \label{eq_part_func_rep}
\mathcal Z_k&[\left\lbrace \mathcal J_a \right\rbrace]= \int \prod_{a=1}^n\mathcal D\Phi_a  \exp \bigg \lbrace - \Delta S_k[\left\lbrace \Phi_a \right\rbrace] \\&-\sum_{a=1}^n \int_{\underline{x}} \big [\frac 12 \left(\partial_{\mu} \Phi_a(\underline{x}) \right) ^2 +U_{B}(\Phi_a(\underline{x})) + \mathcal J_a(\underline x) \Phi_a(\underline x)) \big ] \\& + \frac{\Delta_B}{2}\sum_{a,b=1}^n \int_{\underline{x_1}}\int_{\underline{x_2}}\delta^{(d)}(x_1-x_2) \Phi_a(\underline x_1)\Phi_b(\underline x_2)\bigg \rbrace .
 \end{aligned}
\end{equation}
The $n$-copy action in the above equation is invariant under the $S_n$ permutational symmetry and a global $Z_2$ symmetry as well as the translations and rotations in the Euclidean and Grassmannian subspaces. These symmetries are then shared by a quadratic regulator of the form
\begin{equation}
\Delta S_k=\frac 12  \sum_{a,b=1}^n\int_{\underline{x}}\int_{\underline{x}'} \Phi_a(\underline{x})\mathcal R_{k,ab}(\underline{x},\underline{x}')\Phi_b(\underline{x}'),
\end{equation}
with
\begin{equation}
\label{eq_replicatedregulator}
\mathcal R_{k,ab}(\underline{x},\underline{x}') = \delta_{ab}\delta_{\underline{\theta}\, \underline{\theta}'}\widehat{R}_k(|x-x'|)+\widetilde{R}_k(|x-x'|),
\end{equation}
where $\delta_{\underline{\theta}\, \underline{\theta}'}=(\bar{\theta}-\bar{\theta}')(\theta-\theta')$; $\widehat{R}_k$ and $\widetilde{R}_k$ are IR cutoff functions. The regulator is chosen such that it suppresses the integration over modes with momentum $\vert q \vert \ll k$\cite{berges02,tarjus04} and both functions $\widehat{R}_k$ and $\widetilde{R}_k$ go to zero when $k\to 0$.
In addition to the above mentioned symmetries, $\mathcal Z_k[\left\lbrace \mathcal J_a \right\rbrace]$ in Eq. (5) is invariant under the superrotations when the sources $\hat{J}_a$, $K_a$, $\hat{K}_a$ are set to zero for all copies but one. The theory then reduces to a $1$-copy problem. ($\Delta S_k$ can be made explicitly invariant under the same conditions by choosing $\mathcal R_{k,aa}$ to be a function of the superlaplacian $\Delta_{SS}$ only; as a result, $\widetilde{R}_k(q^2)=-\Delta_B \partial_{q^2}\widehat{R}_k(q^2)$,
where $q$ denotes the momentum in Euclidean space.) The regularization ensures that the modified stochastic field equation has a unique solution at the microscopic scale $\Lambda$ and guarantees that the theory is indeed supersymmetric when $k=\Lambda$.

The central quantity of our NP-FRG approach is the effective average action,\cite{berges02} which is the generating functional of the ``1PI vertices''\cite{zinnjustin89} and is obtained from $\log \mathcal Z_k$ by a (modified) Legendre transform,
\begin{equation}
\Gamma_k[\left\lbrace \Phi_a \right\rbrace] = -\log \mathcal Z_k[\left\lbrace \mathcal J_a \right\rbrace] +\sum_{a=1}^n  \int_{\underline x}  \mathcal J_a(\underline x) \Phi_a(\underline x)-\Delta S_k[\left\lbrace \Phi_a \right\rbrace]
\end{equation}
with $\Phi_a(\underline x)=\delta(\log \mathcal Z_k) / \delta \mathcal J_a(\underline x)$. Its flow with the IR scale $k$ is described by an exact RG equation (ERGE),\cite{berges02}
\begin{equation}
\partial_t \Gamma_k[\left\lbrace \Phi_a \right\rbrace]=\frac 12 \tr  \left\lbrace \partial_t \mathcal R_{k} \;\mathcal P_{k}[\left\lbrace \Phi_a \right\rbrace]\right\rbrace ,
\end{equation}
where $t=\log(k/\Lambda)$ and the trace involves summing over copy indices and integrating over superspace; the modified propagator $\mathcal P_{k,ab}(\underline{x}_1,\underline{x}_2)$ is the (operator) inverse of $\Gamma_k^{(2)} +  \mathcal R_k$ where $\Gamma_k^{(2)}[\left\lbrace \Phi_a \right\rbrace] $ is the second functional derivative of the effective average action with respect to the superfields $\Phi_a(\underline{x})$. (Here, superscripts in parentheses always denote functional derivatives with respect to the field arguments.)

If for each copy a \textit{single} solution of the stochastic field equation is taken into account, it is easily shown that both the random generating functional has a property of  ``Grassmannian ultralocality'', which translates in the fact that both $\log \mathcal Z_k[\left\lbrace \mathcal J_a \right\rbrace]$ and the effective average action $\Gamma_k[\left\lbrace \Phi_a \right\rbrace]$ have a linked expansion in sums over copies and integrals over Grassmann coordinates:
\begin{equation}
\begin{aligned}
\Gamma_k[\left\lbrace \Phi_a \right\rbrace] = \sum_{p\geq1}& \sum_{a_1=1}^n...\sum_{a_p =1}^n
\frac {(-1)^{p-1}}{p!}\times \\& \int_{\underline{\theta}_1}...\int_{\underline{\theta}_p}
\Gamma_{k,p}[\Phi_{a_1}(\underline{\theta}_1),...,\Phi_{a_p}(\underline{\theta}_p)],
\end{aligned}
\end{equation}
where $\Phi_{a}(\underline{\theta})$ denotes a superfield at the Grassmann coordinates $\theta,\bar{\theta}$; (other coordinates are omitted for simplicity) and $\Gamma_{k,p}$ only depends on $p$ superfields at $p$ Grassmannian ``locations'' (on the other hand the dependence on Euclidean coordinates is completely general). The functional $\Gamma_{k,p}[\phi_1,...,\phi_p]$ is related to the $p$th cumulant of the renormalized disorder and, correspondingly, $\Gamma_{k,p}^{(1...1)}[\phi_1,...,\phi_p]$ is related to the $p$th cumulant of the renormalized random field.\cite{tarjus04} By inserting the above formula in Eq. (9), taking derivatives with respect to the superfields, and restricting the superfields to configurations $\Phi_a(\underline{x})=\phi_a(x)$, one obtains a hierarchy of coupled ERGE's for the ``cumulants''  $\Gamma_{k,p}^{(1...1)}[\phi_1,...,\phi_p]$. It is worth stressing that to obtain the flow equation for $\Gamma_{k,p}^{(1...1)}[\phi_1,...,\phi_p]$ with its full functional dependence on the $p$ field arguments, one needs to consider at least $p$ copies. Formally, the whole hierarchy of flow equations for the cumulants can thus be obtained by considering an arbitrary large number of copies. As an illustration, the ERGE for the first cumulant reads
\begin{equation}
\label{eq_flow_Gamma1}
\begin{split}
\partial_t\Gamma_{k,1}\left[\phi_1\right ]=
\dfrac{1}{2} \int_{q} &\bigg \{ \partial_t \widetilde{R}_k(q^2) \widehat{P}_{k; q\,- q}\left[\phi_1\right ]\\& +\partial_t \widehat{R}_k(q^2) \widetilde{P}_{k; q\,- q}\left[\phi_1,\phi_1\right ]\bigg \} ,
\end{split}
\end{equation}
where $\widehat {P}_{k}[\phi ]=\left(\Gamma _{k,1}^{(2)}[ \phi ]+\widehat R_k\right) ^{-1}$ and $
\widetilde {P}_{k}[\phi_1, \phi_2 ]= \widehat {P}_{k}[ \phi_1 ](\Gamma _{k,2}^{(11)}[\phi_1, \phi_2 ]-\widetilde R_k ) \widehat { P}_{k}[ \phi_2 ]$ are obtained as the zeroth-order terms of the expansion of the modified propagator $P_{k;(a,\underline{x}_1)(b,\underline{x}_2)}$ that generalizes Eqs. (7) and (10). The above ERGE coincides with that previously derived without the superfield formalism by means of an expansion in number of free replica sums (when evaluated at $T=0$). The same is true for the ERGE for all higher-order cumulants: for explicit expressions, see [\onlinecite{tarjus04}].

The (super)symmetries of the modified action in Eq. (5) are linearly
realized and induce a set of WT identities for the 1PI
generating functional $\Gamma_{k}$.\cite{zinnjustin89} 
%For the
%superrotations (with the restriction to $1$ copy via the proper choice
%of sources), one has
%\begin{equation}
%\int_{\underline{x}}\Phi(\underline{x})\mathcal Q_{\mu}\Gamma_{k;\underline{x}}^{(1)}[\Phi]=0
%\end{equation}
%and similarly with $\bar{\mathcal Q}_\mu$. 
Taking functional derivatives of these identities with respect to the
superfield and evaluating the resulting relations for superfield
configurations $\Phi(\underline{x})=\phi(x)$ leads to relations for
the cumulants. The most powerful relations mix cumulants of orders $p$
and $(p+1)$, the first nontrivial illustration of the latter kind
being
\begin{equation}
\begin{aligned}
\partial_{1\mu}\Gamma_{k2;x_1;x_2}^{(11)}[\phi,\phi] &- \frac{\Delta_B}{2} (x_1^\mu-x_2^\mu)\Gamma_{k1;x_1,x_2}^{(2)}[\phi]  = \\&- \int_{x_3}\phi(x_3)
\partial_{3\mu}\Gamma_{k2;x_1,x_3;x_2}^{(21)}[\phi,\phi],
\end{aligned}
\end{equation}
which for fields that are also uniform in the Euclidean space gives a relation similar to that for the cutoff functions: $\Gamma_{k,2}^{(11)}(q^2;\phi,\phi) = \Delta_B \partial_{q^2}\Gamma_{k,1}^{(2)}(q^2;\phi)$.

An important feature of the present superfield theory is that SUSY leads to DR: this is obtained \textit{nonperturbatively} by combining the WT identity in Eq.~(12) with the ERGE for the first cumulant in Eq.~(11) and by following the line of reasoning of Refs. [\onlinecite{cardy83}]. As one knows that DR does not hold in low enough dimension, what then goes wrong in the formalism ? The answer is that SUSY, more precisely invariance under the superrotations when the theory is restricted to a single copy, is spontaneously broken along the flow and that a singularity occurs. From an analysis of the structure of the flow equations, we expect that breaking of DR requires the presence of a linear  ``cusp''  in the field dependence of $\Gamma_{k2}^{(11)}$, cusp that should appear \textit{at a finite scale} during the RG flow. (On the other hand, weaker nonanalyticities in $\Gamma_{k2}^{(11)}$ and nonanalyticities in higher-order cumulants can only appear at the fixed point, in the limit $k\rightarrow0$, thereby preserving the DR property.) This of course must be checked in actual calculations, which is what we provide below.

If SUSY is spontaneously broken, how can one continue the RG flow for
the effective average action ? The original formal construction
\textit{a priori} loses its meaning, but a \textit{nontrivial
  continuation} can be found if (i) one assumes that, except for the
superrotations, all of the properties and symmetries of $\Gamma_{k}$
remain valid; most importantly, this includes the ``Grassmannian
ultralocality'' encompassed in Eq.~(10) that enforces single-solution
dominance\cite{tissier10}, (ii) one only considers ERGE's for
cumulants evaluated for generic (nonequal) field arguments, so that a
putative nonanalytic dependence can freely emerge, and (iii) one
modifies the regulator by replacing $\Delta_B$ by a running $\Delta_k$
which is the typical strength of the renormalized random field at
scale $k$. More specifically, and in order to reach a fixed point and
a scale-free solution describing the critical behavior of the RFIM, we
choose $\widehat{R}_k(q^2)=Z_k k^2 r(q^2/k^2)$ and
$\widetilde{R}_k(q^2)=-(\Delta_k/Z_k) r'(q^2/k^2) $; $Z_k$ and
$\Delta_k$ are respectively obtained from
$\partial_{q^2}\Gamma_{k1}^{(2)}(q^2)$ and $\Gamma_{k2}^{(11)}(q^2)$
evaluated for $q^2=0$ at zero field, and choices for the function
$r$ are given in [\onlinecite{berges02,tarjus04,delamotte}]. From
Eq.~(12) and below, one can see that so long as SUSY is not broken,
$\Delta_k=\Delta_B Z_k$ and the regulator is SUSY invariant, which
guarantees the consistency of the RG description.

\begin{figure}[t]
\includegraphics[width=1.\linewidth]{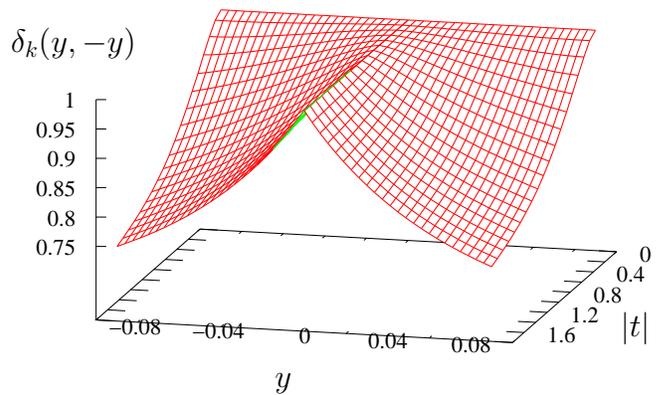}
\caption{\label{fig_larkin_y} NP-FRG flow of the dimensionless cumulant $\delta_k(\varphi+y,\varphi-y)$ in $d=4<d_{DR}$ for $\varphi=0$ and for initial conditions close to the critical point. A linear cusp in $\vert y\vert$ appears at a finite RG ''time'' $\vert t\vert=\log(\Lambda/k)$.}
\end{figure}
Finally, we provide a SUSY-compatible nonperturbative approximation
scheme for the ERGE. We combine truncations in the derivative
expansion, which approximate the long-distance behavior of the 1PI
vertices, and in the expansion in cumulants of the renormalized
disorder. The WT identities require that the orders of truncation in
the two types of expansions be related. The minimal truncation that
can already describe the long-distance physics of the RFIM and does
not {\em explicitly} break SUSY is the following:
\begin{equation}
\begin{aligned}
&\Gamma_{k,1}[\phi]=\int_{x}\left[ U_k(\phi(x))+\frac{1}{2}Z_k(\phi(x))(\partial_{\mu}\phi(x))^2 \right],\\&
\Gamma_{k,2}[\phi_1,\phi_2]=\int_{x}V_k(\phi_1(x),\phi_2(x)),
\end{aligned}
\end{equation}
with the higher-order cumulants set to zero. Inserted in the ERGE for
the cumulants, the above ansatz provides $3$ coupled flow equations
for the $1$-copy potential $U_k(\phi)$ that describes the
thermodynamics of the system, the field renormalization function
$Z_k(\phi)$, and the $2$-copy potential $V_k(\phi_1,\phi_2)$ from
which one obtains the second cumulant of the renormalized random
field, $\Gamma_{k2}^{(11)}(q^2=0;\phi_1,\phi_2) \equiv
\Delta_k(\phi_1,\phi_2) =\partial _{\phi_1}\partial
_{\phi_2}V_k(\phi_1,\phi_2)$.
\begin{figure}[t]
\includegraphics[width=1.\linewidth]{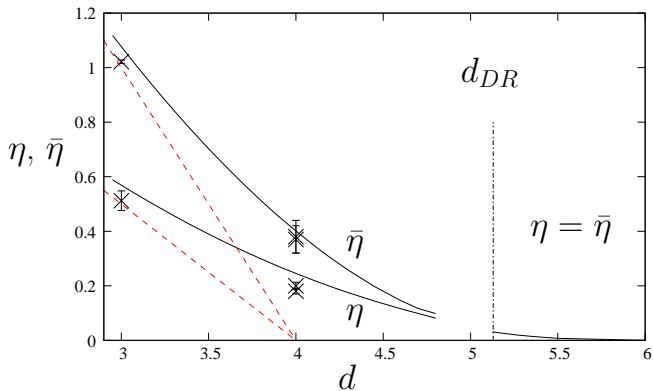}
\caption{Anomalous dimensions $\eta$ and $\bar{\eta}$ versus $d$. DR is observed above $d_{DR}\simeq 5.1$. $\eta$ and $\bar\eta$ satisfy the required upper ($\bar \eta \leq 2\eta$) and lower bounds (red dashed lines).\cite{nattermann98} Stars correspond to simulation results.\cite{middleton,hartmann02} The   region just below $d_{DR}$ is unfortunately numerically difficult to access.}
\end{figure}

To search for the fixed point that controls the critical behavior
(associated with a spontaneous breaking of the $Z_2$ symmetry), the
flow equations must be recast in a scaled form. The fixed point being
a zero-temperature one,\cite{villain84,nattermann98} the spatial decay
of the correlations [see below Eq. (11)] at criticality is now characterized by two ''anomalous dimensions'' $\eta$ and $\bar \eta$,
\begin{equation}
\hat P(r)\sim r^{-(d-2+\eta)},\qquad \tilde P(r)\sim r^{-(d-4+\bar\eta)},
\end{equation}
with $\eta\leq \bar\eta\leq2\eta$, and one has to introduce 
scaling dimensions involving an additional critical
exponent.\cite{tarjus04} The resulting equations are generalizations
of those shown in Ref.~[\onlinecite{tarjus04}] and are not displayed
here. We have solved these coupled partial differential equations
numerically, looking for the proper (critical) fixed point as a
function of dimension (more details will be given elsewhere). This
procedure is numerically very demanding and requires handling $3$
coupled equations for $2$ functions of $1$ variable ($U_k$ and $Z_k$)
and $1$ function of $2$ variables ($\Delta_k$).

An important property of the present theory is that if in the limit
$\phi_2 \rightarrow \phi_1$,
$\Delta_k(\phi_1,\phi_2)=\Delta_{k0}(\phi) +\Delta_{k2}(\phi)(\phi_1 -
\phi_2)^2 +\cdots$ with $\phi=(\phi_1 + \phi_2)/2$, then the flow of
$\Delta_{k0}(\phi)$ coincides with that of $Z_k(\phi)$: this is
precisely the WT relation derived from Eq. (13), and DR \textit{exactly} follows. On
the other hand, a spontaneous breaking of the SUSY and of the
associated WT identity occurs whenever $\Delta_{k2}(\phi)$ diverges
and $\Delta_k$ has a cusp-like singularity in the form
$\Delta_k(\phi_1,\phi_2)=\Delta_{k0}(\phi) +\Delta_{ka}(\phi)|\phi_1 -
\phi_2| +\cdots$ as $\phi_2 \rightarrow \phi_1$.

We find that the solution without cusp is stable and that $\eta(d)=\bar\eta(d)=\eta^{Ising}(d-2)$,
in agreement with the DR prediction, above a critical dimension
$d_{DR} \simeq 5.1$. For $d<d_{DR}$, we obtain a once unstable ``cuspy''
fixed point (see Fig. 1) and DR is broken: the exponents $\eta$ and $\bar\eta$
bifurcate, with $\eta(d)< \bar\eta(d)$ (see Fig. 2). In $d=3$, we find
$\eta \simeq 0.57, \bar{\eta}\simeq 1.08$ and in $d=4$, $\eta \simeq
0.24, \bar{\eta}\simeq 0.40$: this is in good agreement with the existing
estimates,\cite{hartmann02,middleton} which gives support to the whole scenario (the results are also $1$-loop exact near $d=6$). In addition, the continuous
variation of $\eta$ and $\bar\eta$ with $d$ and the existence of a
critical dimension above which $\eta(d)=\bar\eta(d)$ contradicts the
claim that the two exponents are always related by a fixed ratio
$\bar\eta(d)=2 \eta(d)$.\cite{schwartz85b}

In conclusion, the present study provides key pieces for a complete
resolution of the long-standing puzzles associated with the critical
behavior of the RFIM. In doing so, we have developed tools that may prove useful in other contexts where the need to select a unique solution of a stochastic field equation arises, as in ''glassy'' systems, turbulence or nonabelian gauge field theories.

\end{document}